# Image Reconstruction with $B_0$ Inhomogeneity using an Interpretable Deep Unrolled Network on an Open-bore MRI-Linac

Shanshan Shan, Yang Gao, David E. J. Waddington, Hongli Chen, Brendan Whelan, Paul Z. Y. Liu, Yaohui Wang, Chunyi Liu, Hongping Gan, Mingyuan Gao, Feng Liu

*Abstract*—MRI-Linac systems require fast image reconstruction with high geometric fidelity to localize and track tumours for radiotherapy treatments. However, $B_0$ field inhomogeneity distortions and slow MR acquisition potentially limit the quality of the image guidance and tumour treatments. In this study, we develop an interpretable unrolled network, referred to as RebinNet, to reconstruct distortion-free images from B0 inhomogeneity-corrupted k-space for fast MRI-guided radiotherapy applications. RebinNet includes convolutional neural network (CNN) blocks to perform image regularizations and nonuniform fast Fourier Transform (NUFFT) modules to incorporate B0 inhomogeneity information. The RebinNet was trained on a publicly available MR dataset from eleven healthy volunteers for both fully sampled and subsampled acquisitions. Grid phantom and human brain images acquired from an open-bore 1T MRI-Linac scanner were used to evaluate the performance of the proposed network. The RebinNet was compared with the conventional regularization algorithm and our recently developed UnUNet method in terms of root mean squared error (RMSE), structural similarity (SSIM), residual distortions, and computation time. Imaging results demonstrated that the RebinNet reconstructed images with lowest RMSE (<0.05) and highest SSIM (>0.92) at four-time acceleration for simulated brain images. The RebinNet could better preserve structural details and substantially improve the computational efficiency (ten-fold faster) compared to the conventional regularization methods, and had better generalization ability than the UnUNet method. The proposed RebinNet can achieve rapid image reconstruction and overcome the $B_0$ inhomogeneity distortions simultaneously, which would facilitate accurate and fast image guidance in radiotherapy treatments.

## I. INTRODUCTION

MRI-guided radiotherapy (MRgRT) workflow has recently attracted increasing attention for accurate tumour treatments due to its excellent soft-tissue contrast and lack of additional ionizing radiation exposure compared to other imaging techniques such as computed tomography (CT) [1, 2]. Hybrid systems integrating an MRI scanner and a medical linear accelerator (Linac) have been developed to enable fast tumour tracking and image guidance, which would facilitate optimized dose delivery and adaptive patient-specific treatment strategies [3, 4]. Such MRI-Linac systems show great promise for the translation of accurate image-guided radiotherapy treatments in routine clinical practice [5].

In a conventional MR image reconstruction system, the main magnetic ($B_0$) field is assumed to be uniform in the field of view (FOV) [6-8] for the MR signal spatial encoding. However, in practice, generating a perfectly uniform $B_0$ field is not always achievable due to hardware constraints on main magnet design and manufacture. Sometimes, the patient "friendliness" design of the magnet, such as short and wide-bore systems leads to trade-offs in $B_0$ field homogeneity [9]. In particular, the Australian 1T MRI-Linac scanner applies an open-bore/split magnet with a large central gap (50cm) to facilitate patient positioning and radiation dose delivery [10, 11]. The split bore configuration is prone to restrict $B_0$ field homogeneity, which will undermine the spatial encoding and thus produce geometric distortions in reconstructed MR images [12]. The geometric distortions cause target position errors and potentially degrade precise radiation delivery in the MRI-Linac tumour treatments [13, 14]. This effect will be particularly problematic for MRgRT applications where high geometric fidelity is required [15, 16].

This work was supported by the National Natural Science Foundation of China under Grant No. 62301352 and 62301616, and the Australian Government National Health and Medical Research Council (NHMRC) Program (1132471).

Shanshan Shan is with Center for Molecular Imaging and Nuclear Medicine, State Key Laboratory of Radiation Medicine and Protection, School for Radiological and Interdisciplinary Sciences (RAD-X), Soochow University, Suzhou 215123, China, also with Image X Institute, Sydney School of Health Sciences, Faculty of Medicine and Health, University of Sydney, Sydney 2015, Australia, and also with Department of Medical Physics, Ingham Institute of Applied Medical Research, Liverpool, 2170, Australia (ssshan@suda.edu.cn).

Yang Gao is with School of Computer Science and Engineering, Central South University, Changsha 410083, China, also with School of Electrical Engineering and Computer Science, University of Queensland, Brisbane 4072, Australia (yang.gao@csu.edu.cn).

David E. J. Waddington, Brendan Whelan and Paul Z.Y. Liu are with Image X Institute, Sydney School of Health Sciences, Faculty of Medicine and Health, University of Sydney, Sydney 2015, Australia, and also with Department of Medical Physics, Ingham Institute of Applied Medical Research, Liverpool, 2170, Australia (david.waddington@sydney.edu.au).

Hongli Chen and Feng Liu are with School of Electrical Engineering and Computer Science, University of Queensland, Brisbane 4072, Australia (feng@eecs.uq.edu.au).

Yaohui Wang is with Institute of Electrical Engineering, Chinese Academy of Science, Beijing 100190, China (yhw@mail.iee.ac.cn).

Chunyi Liu and Mingyuan Gao are with Center for Molecular Imaging and Nuclear Medicine, State Key Laboratory of Radiation Medicine and Protection, School for Radiological and Interdisciplinary Sciences (RAD-X), Soochow University, Suzhou 215123, China (gaomy@suda.edu.cn).

Hongping Gan is with School of Software, Northwestern Polytechnical University, Taicang 215400, China (ganhongping@nwpu.edu.cn).

Given the knowledge of $B_0$ field inhomogeneity, image distortions can be retrospectively corrected by post-processing interpolation algorithms [17]. However, such image-domain methods can result in image blurring and resolution loss due to the intrinsic smoothing effect of interpolation operations, particularly at the edges of large FOVs [18, 19]. Alternatively, geometric deformations can be corrected prospectively during the image reconstruction process [20]. This method incorporates distortion information into the gradient encoding matrix, translating the sophisticated image reconstruction under field deviations into an ill-conditioned and nonlinear inverse problem. Iterative minimization algorithms with regularizations such as wavelet transform [21], low rank [22], and total variation [23] are normally applied to solve the ill-conditioned problem and to reconstruct distortion-free images directly from the raw k-space domain. This k-space domain method can effectively overcome the smoothing effect of interpolation-based methods [24]. However, it is non-trivial to select optimal regularization parameters, and the iterative algorithms are computationally expensive, which is challenging for fast image guidance in radiotherapy treatments [25].

Deep learning has shown promising performances in solving MR regularization problems, enabling accurate and fast online image reconstruction [26-29]. Convolutional neural networks (CNN)-based methods have been widely investigated to reconstruct high-quality MRI images from incomplete k-space or aliased images [30]. A 2D UNet and a modified GAN were trained to recover distortion-free images with super-resolution for the single-shot echo planar imaging (EPI) [31, 32]. Recently, we proposed an image-domain CNN-based method (UnUNet) and a physics-guided network (DCReconNet) to rapidly reconstruct images from subsampled k-space in the presence of gradient nonlinearity (GNL) [28, 33]. In this work, we develop an interpretable unrolled network (RebinNet) to reconstruct images with corrected $B_0$ inhomogeneity distortions for MRgRT applications on an MRI-Linac. The RebinNet included CNN blocks to perform image regularizations and Nonuniform Fast Fourier Transform (NUFFT) to operate $B_0$ inhomogeneity spatial encoding. Grid phantom measurements with reversed gradient polarities were used to measure the $B_0$ inhomogeneity information. Subsampled k-space was integrated into the RebinNet architecture to further reduce the acquisition time. The proposed RebinNet was trained on a publicly available MR brain dataset with 11 healthy volunteers, and then evaluated on grid phantom and volunteer brain images acquired from an open-bore 1T MRI-Linac scanner with fully sampled and subsampled acquisitions. The RebinNet was compared with the conventional regularization algorithm and the image-domain UnUNet method in terms of image quality and computational cost.

## II. METHODS AND MATERIALS

*A. $B_0$ inhomogeneity spatial encoding and image reconstruction*

Given the theoretical/undistorted location $L$, the distorted location $\tilde{L}^+$ with a positive encoding gradient and the location $\tilde{L}^-$ with a negative encoding gradient are governed by the equations below [9]:

$$\tilde{L}^+ = L + \frac{dB_G(L)}{G} + \frac{dB_0(L)}{G} \quad (1)$$

$$\tilde{L}^- = L + \frac{dB_G(L)}{G} - \frac{dB_0(L)}{G} \quad (2)$$

where $dB_G(L)$ and $dB_0(L)$ denote gradient field and $B_0$ field perturbations at the location of $L$, respectively. $G$ represents the applied gradient strength. $\frac{dB_G(L)}{G}$ and $\frac{dB_0(L)}{G}$ represent the GNL- and $B_0$ inhomogeneity-induced distortions, respectively. The sign of $B_0$ inhomogeneity distortion $\frac{dB_0(L)}{G}$ is affected by the gradient polarity and thus the distortion can be calculated by averaging the displacements:

$$\frac{dB_0(L)}{G} = \frac{\tilde{L}^+ - \tilde{L}^-}{2} \quad (3)$$

Eq. (3) shows that $B_0$ inhomogeneity effect is inversely proportional to the applied gradient strength, which can be calculated by the following equation:

$$G = \frac{W*S}{\gamma*V} \quad (4)$$

where $W$ is the imaging bandwidth and $S$ is the image size; $\gamma$ and $V$ represent the gyromagnetic ratio and the size of FOV, respectively.

The forward spatial encoding process with $B_0$ field inhomogeneity can be formulated as:

$$m(E_{B_0}°F)x = b \quad (5)$$

where $b$ is the measured k-space data; $m$ represents the undersampling mask and $x$ is the distortion-free image; $F$ denotes the theoretical Fourier Transform operator with the kernel of $e_{k,L} = e^{-2\pi jkL}$; $E_{B_0}$ is the distorted encoding operator caused by $B_0$ field inhomogeneity in the form of $e_{k,\Delta(L)} = e^{-2\pi jk\Delta(L)}$ and $\Delta(L) = \frac{dB_0(L)}{G}$ is the $B_0$ inhomogeneity displacement at location $L$. $E_{B_0}°F$ represents the element-wise multiplication of matrix $E_{B_0}$ and $F$. The distortion-free image can then be reconstructed by solving the minimization problem:

$$x = \underset{x}{\mathrm{argmin}}\{\|m(E_{B_0}°F)x - b\|_2^2 + \lambda P(x)\} \quad (6)$$

where $\|m(E_{B_0}°F)x - b\|_2^2$ is the L2-norm operation which promotes the data fidelity between the measured and estimated data. $P(x)$ represents the sparsity regularization term (e.g., wavelet, low rank, and total variation) with weighting parameter $\lambda$ [34, 35]. It is noted that $E_{B_0}°F$ denotes a nonuniform to uniform spatial mapping which can be implemented by the type-I NUFFT algorithm [36]. Conventional regularization methods based on iterative algorithms [37] have been typically used to solve the above minimization problem in Eq. (6). However, it requires a cumbersome fine-tuning process to determine the optimal weighting parameter $\lambda$, and the iterative algorithms are computationally expensive, making them impractical for the clinical translation.

*B. Network architecture*

Recently, model-driven unrolled neural networks have been increasingly used for solving MR inverse problems, which incorporate MR physical models and have well-defined interpretability [38]. Inspired by the unrolled network architecture, we developed a $B_0$ inhomogeneity distortion-free reconstruction pipeline to solve the ill-conditioned problem described by Eq. (5) through the following equation:

$$x = \underset{x}{\mathrm{argmin}}\{\|m(E_{B_0}°F)x - b\|_2^2 + CNN(x)\} \quad (7)$$

where $CNN(x)$ represents convolutional neural networks that are used to learn effective regularizations. As shown in Fig. 1(a), the input of the RebinNet includes the undersampled k-space data and the $B_0$ inhomogeneity displacement information, and the output is the reconstructed distortion-free image. The proposed network consists of seven iterative soft shrinkage-thresholding blocks, and each block includes a data fidelity calculation and a *CNN*-based latent regularization (see Fig. 1(b)). The forward and backward transforms with a soft-thresholding operator are implemented to reduce image artifacts. Two linear convolutional operators and a rectified linear unit (ReLU) are applied to form each transform operation. A residual block is added to further facilitate the network performance.

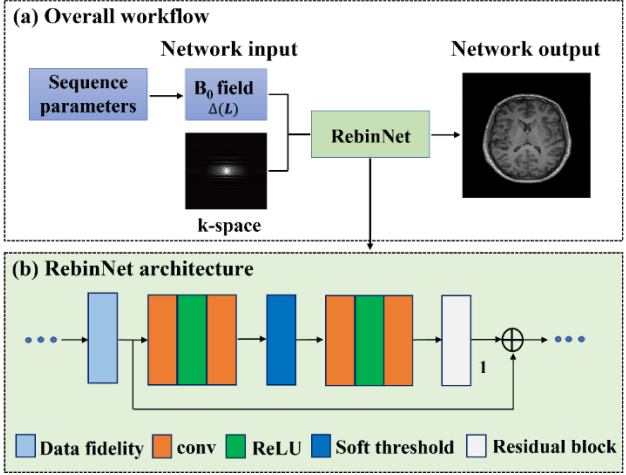

Fig. 1. The overall workflow of the proposed method (a) and the RebinNet architecture (b). It takes both the distorted k-space data and the $B_0$ inhomogeneity displacement as the input, and outputs the distortion-free image.

*C. Data preparation for network training and testing*

In this work, a public T1-weighted brain image dataset from 11 healthy volunteers was used to generate training data for the RebinNet. The brain images were acquired from a whole-body MRI system (Magnetom Tim Trio; Siemens Healthcare, Germany) and the imaging parameters are: image size=320 × 320 × 256, resolution=0.7×0.7×0.7 mm³ isotropic, TE/TR=2.13 ms/2.4 s, echo spacing=65 ms, and phase encoding direction: R/L. The brain images were uniformly located in the entire region of interest (ROI) of 30×30×30 cm³ to simulate distorted k-space data. 3000 brain images from ten healthy volunteers were used as training data and 300 images from the eleventh volunteer were selected to generate the testing data. As shown in Fig. 2, a specially designed 3D phantom [39] was scanned from a body coil on the Australian MRI-Linac scanner with reversed gradient polarities to measure the $B_0$ field inhomogeneity information on phantom marker positions. The imaging parameters were image size=130×110 × 192, image resolution=1.8 mm × 2 mm × 1.8 mm, pixel bandwidth=202 Hz, turbo spin echo sequence (TSE), TE/TR=15 ms/5.1 s and phase encoding direction: R/L. Our previously developed spherical harmonic (SH) method [40] was used to calculate the $B_0$ inhomogeneity displacement at any position in the FOV based on the phantom measurements. The calculated $B_0$ field information was then integrated into the spatial encoding process in Eq. (5) to simulate distortion-corrupted k-space data. These simulated k-space data were retrospectively subsampled by 1D random undersampling masks at acceleration factors (AFs) of 2, 4, and 6 along the phase encoding direction.

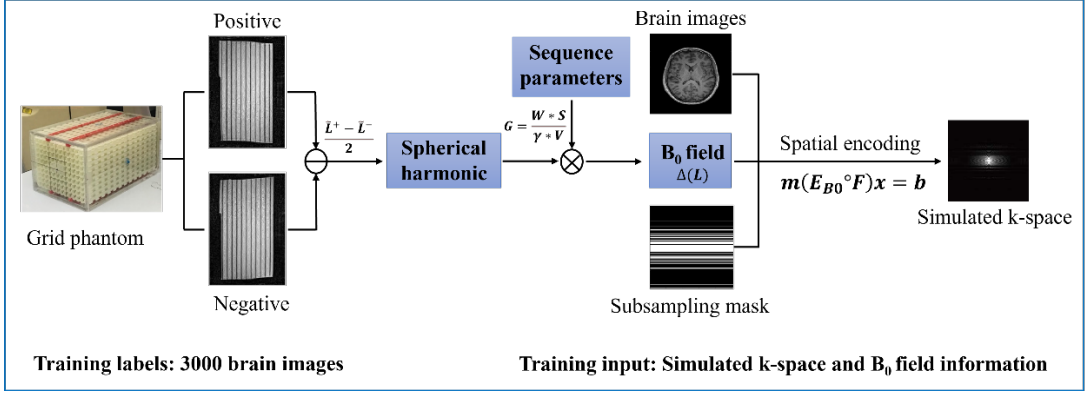

Fig. 2. The training data preparation pipeline, consisting of phantom measurements, spherical harmonic calculation and $B_0$ inhomogeneity spatial encoding.

To test the performance of our network, another two scans of the 3D distortion phantom were conducted with 6cm-shifted phantom isocenter along y direction, and reversed gradient polarities with full sampling were applied for each scan. The pixel bandwidth for both scans was 101 Hz and 202 Hz, respectively and the other imaging parameters were image size=130×110×192, image resolution=1.8 mm×2 mm×1.8 mm, turbo spin echo sequence (TSE), TE/TR=15 ms/5.1 s and phase encoding direction: R/L. The fully sampled phantom data were then retrospectively subsampled with AF=4. A healthy volunteer's brain was scanned with a six-channel head coil with the following parameters: image size=256 × 256 × 12, image resolution=0.98 mm×0.98 mm×5 mm, pixel bandwidth=203 Hz, TSE sequence, TE/TR=77 ms/8 s and phase encoding direction: R/L. The acquisitions were retrospectively subsampled with AFs=2 and 4 using 1D random undersampling mask. For multi-channel imaging, the sum-of-squares (SoS) operation was used to combine all channel data so that the coil sensitivity maps were not required for the network reconstruction.

*D. Evaluation*

Mean squared error (MSE) loss was adopted for the network training using the Adam optimizer [41] with a mini-batch size of 32. The network was trained for 100 epochs and the learning rate was 0.001. The training took approximately 20 hours on a high-performance computer equipped with a Nvidia Tesla V100 P32 GPU. All human studies were conducted with the approval of the Institutional Review Board (IRB).

The conventional regularization-based compressed sensing (CS) method with the NUFFT operation was implemented to solve the minimization problem in Eq. (6) and to reconstruct the distortion-free images, referred to as DFCS. A grid search was used to determine the optimal regularization parameters that minimized the MSE between the reconstructed and reference images. The distortion-free zero-filling reconstruction method was referred to as DFZF. An additional image-domain UnUNet method was also implemented to solve Eq. (6), where a standard unrolled network was used to reconstruct distorted images from the undersampled k-space data and a ResUNet was then applied to correct geometric distortion from the image domain. The UnUNet method was trained on the same dataset as described in Section 2.3. The RebinNet was compared with DFZF, DFCS, and UnUNet methods at AFs=2, 4, and 6. The conventional Fourier Transform reconstruction and the RebinNet were also implemented for fully sampled acquisitions, referred to as FT and RebinNet-FS. The root mean squared error (RMSE) and structural similarity index (SSIM) metrics were calculated to quantitatively evaluate the image reconstruction quality. 3289 marker positions extracted from the specially designed 3D phantom were used to quantitatively measure the geometric distortion on reconstructed images.

## III. RESULTS

*A. Simulation results*

Fully sampled brain images with $B_0$ inhomogeneity were simulated at different pixel bandwidths for positive and negative gradient encoding polarities to validate the effectiveness of the proposed RebinNet for distortion correction, as shown in Fig. 3. Yellow lines are the contours of ground truth images. The image contours of the conventional FT reconstructions do not match with yellow lines due to $B_0$ inhomogeneity-induced distortions. As indicated by the red arrows, image distortions increase inversely with the bandwidth. When applying the reversed gradient polarity, different geometric displacement patterns were observed on FT-reconstructed images. By contrast, the RebinNet dramatically reduced distortions and negligible errors were observed in RebinNet results, as illustrated by the error maps.

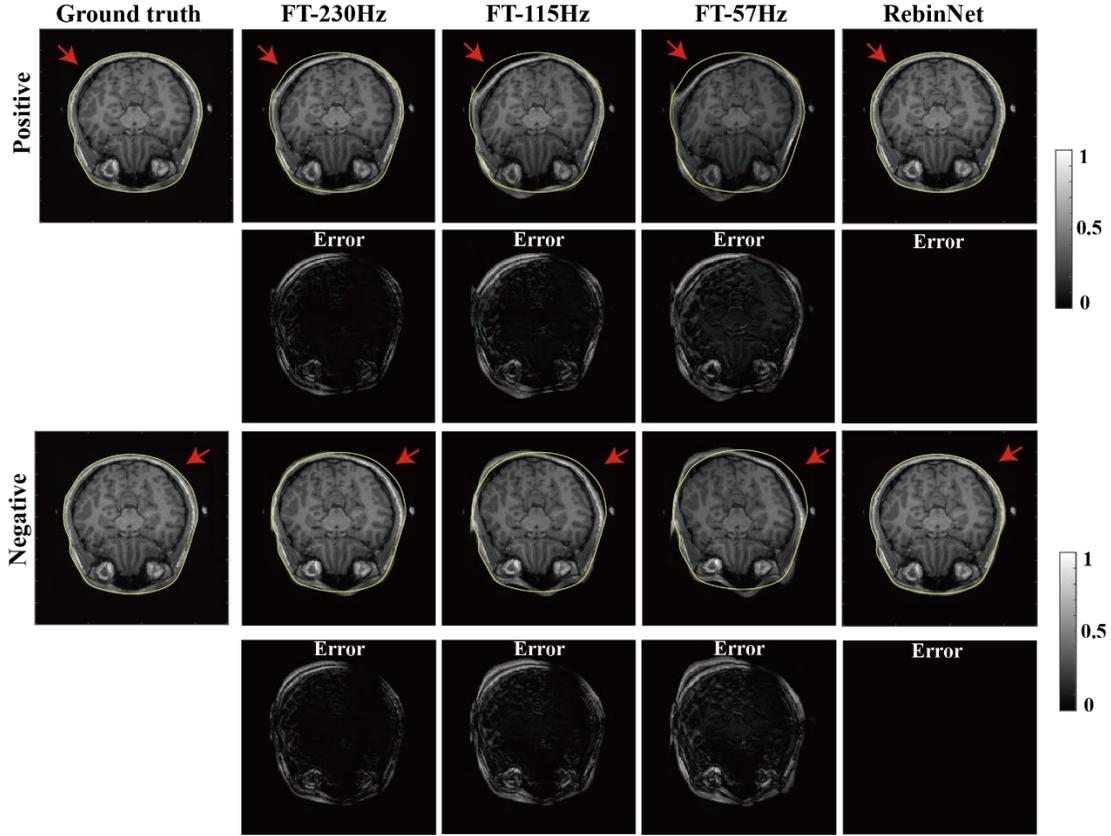

Fig. 3. Fully sampled brain images with positive and negative gradient polarities. The bandwidth frequency of 230Hz, 115Hz and 57Hz was used to simulate the k-space data that was then reconstructed by the conventional FT method. The simulated k-space data at 230Hz was reconstructed by the RebinNet method for comparison. Error maps between reconstructed images and ground truth were shown at the bottom.

Subsampled data at AF = 2, 4, 6 were also simulated to compare the performances of the proposed RebinNet with DFZF, DFCS, and UnUNet, and the results are demonstrated in Fig. 4. As indicated by the yellow lines, these four methods successfully correct geometric distortions compared with the simple FT method. Image artifacts and noise in DFZF results were considerably alleviated by DFCS, UnUNet and the proposed RebinNet. These three algorithms reconstructed images with comparable image quality at AF=2. The DFCS method reconstructed images with signal blurring and detail loss at AF of 4 and 6, as pointed out by the yellow arrows. The image structural details were well preserved on UnUNet and RebinNet images, demonstrating that the neural network approach can achieve superior performance for higher AF subsampling cases. Quantitatively, the UnUNet and RebinNet resulted in lower RMSE and higher SSIM than the DFCS method.

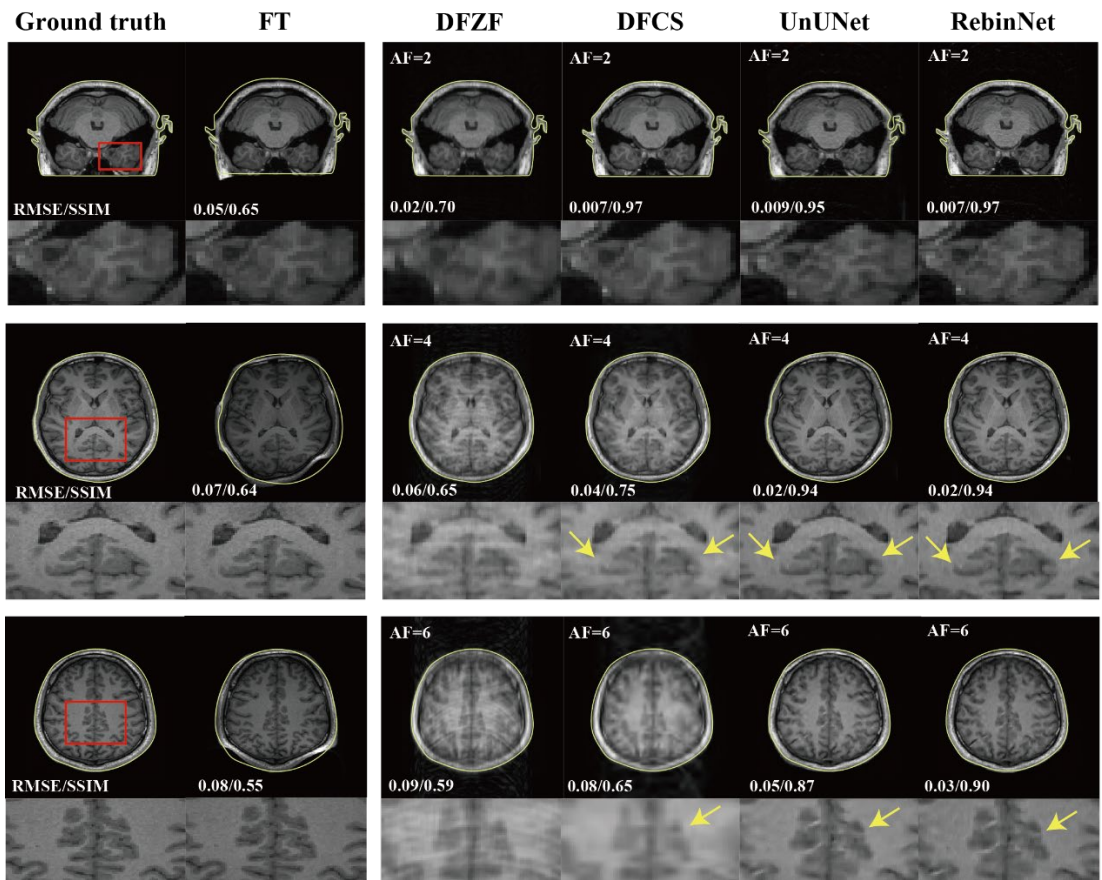

Fig. 4. Brain image reconstructions from distortion-corrupted k-space data at acceleration factors of 2, 4 and 6. DFZF, DFCS, UnUNet and RebinNet reconstructions were compared for axial slices at z= 36mm (AF=2), z=132mm (AF=4) and z=148mm (AF=6), respectively. Fully sampled data was reconstructed by the conventional FT method. Zoomed regions (red rectangle) were shown at the bottom of each image. Yellow lines represent the contours of ground truth brain images. Yellow arrows indicate that the UnUNet and RebinNet methods can preserve better structural details than the other methods.

Comparison of RMSE and SSIM metrics for different reconstruction methods on the 300 testing brain images are shown in Fig. 5. The median and maximum RMSE values for the DFZF-reconstructed images are 0.05 and 0.09, respectively; whereas these two values are lower than 0.04 and 0.06 for the other three methods. It is noticeable that the RebinNet and UnUNet methods provided lower RMSE values than the DFCS method, demonstrating more accurate image reconstructions. In terms of the SSIM level, the images reconstructed by the two neural network approaches show the highest median value (~0.95) than those reconstructed by DFZF (0.68) and DFCS (0.79) methods.

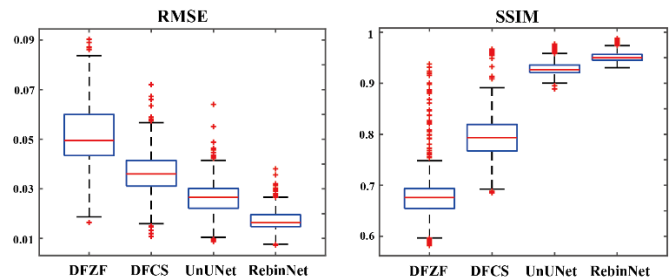

Fig. 5. Boxplots of RMSE (left) and SSIM (right) values across 300 testing brain images reconstructed by DFZF, DFCS, UnUNet and RebinNet methods at AF=4. The minimum, first quartile (25%), median (50%), third quartile (75%) and the maximum values were statistically analysed. Red crosses represent outliers that account for 0.7% of total samples.

*B. Experimental results*

  *1) Phantom results*: A 3D grid phantom was scanned from a body coil on an MRI-Linac scanner with pixel bandwidth = 202 Hz to evaluate the performance of the proposed method. As shown in Fig. 6, the fully sampled FT-reconstructed images show opposite geometric distortion patterns for positive and negative gradient polarities, demonstrating that the $B_0$ inhomogeneity effect is dependent on the sequence parameters. The image distortions were substantially reduced on RebinNet-FS images, which were used as references for subsampled acquisitions at AF of 4. A comparison of different methods on this phantom at AF=4 was also shown in the right part of Fig. 6. As indicated by the yellow arrows, undesired image blurring and artifacts in UnUNet reconstructions were reduced by the RebinNet. The phantom images acquired with pixel bandwidth = 101 Hz were shown in Supporting Information Figure SI, and the RebinNet gave the best image quality compared with the other methods, which is consistent with the results in Fig. 6.

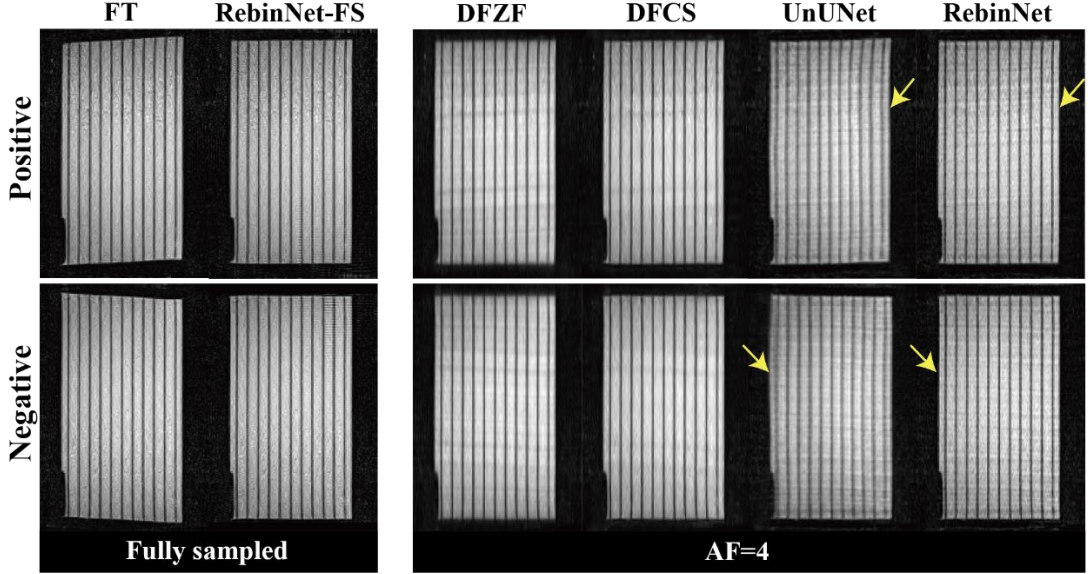

Fig. 6. Grid phantom images acquired with positive (the top row) and negative (the bottom row) gradient polarities on the MRI-Linac system with pixel bandwidth = 202 Hz at the location of x=-33.7 mm. Fully sampled data was reconstructed by the FT and RebinNet-FS methods. Subsampled data at AF=4 was reconstructed by DFZF, DFCS, UnUNet and RebinNet methods, respectively.

3289 phantom markers were extracted from the grid phantom images before and after RebinNet corrections to quantitatively measure the geometric displacements. The distribution of image distortions on 3289 markers in the whole FOV is shown in Fig. 7. Before correction, geometric distortions on 2539 markers are less than 2 mm and 383 markers have distortions over than 4 mm. After applying the RebinNet, the residual displacements of 3278 markers are within 1mm and only 11 markers' distortions are larger than 1mm. The maximal displacement of uncorrected images is 12.2 mm; while the RebinNet reduced it within 2mm, showing a significant improvement on geometric fidelity. Similarly, the RMSE of corrected markers is only 0.16 mm, which is smaller than one-tenth of uncorrected ones (2.8 mm).

  *2) In vivo brain results*: Fig. 8 compares the proposed RebinNet with DFZF and DFCS for the acquired *in vivo* brain data at AF = 2 and 4. Fully sampled images reconstructed by the RebinNet-FS were used as references. Comparable image quality with similar RMSE (0.01) and SSIM (0.94) values was achieved for DFCS and RebinNet methods at AF=2. Image structural details for AF=4 were better preserved in the RebinNet than DFCS, which is consistent with the results of Fig. 4 and Fig. 6.

*C. Computational efficiency*

The DFCS, UnUNet and RebinNet algorithms were implemented on a desktop computer equipped with an Intel Xeon central processing unit (CPU) @ 3.7 GHz and RAM (16 GB) on a Windows 10 Enterprise system. The inference time of DFCS method on an image size of 256×256 was 30s, while the UnUNet and RebinNet methods required 3s, showing a ten-fold reduction in computational cost. In addition, we executed the two neural network approaches on a high-performance computer equipped with an Nvidia Tesla V100 P32 GPU and the latency was only 300 ms, demonstrating the fast image reconstruction potential for MRgRT applications.

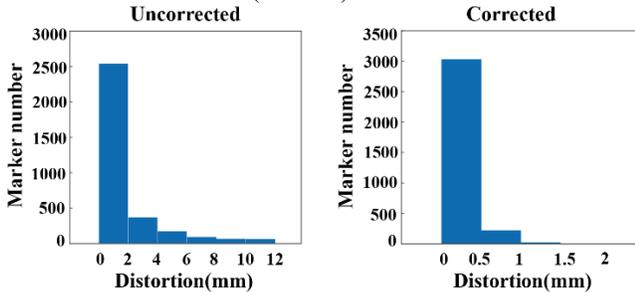

Fig. 7. Geometric distortion distribution on 3289 markers in the FOV before and after correction.

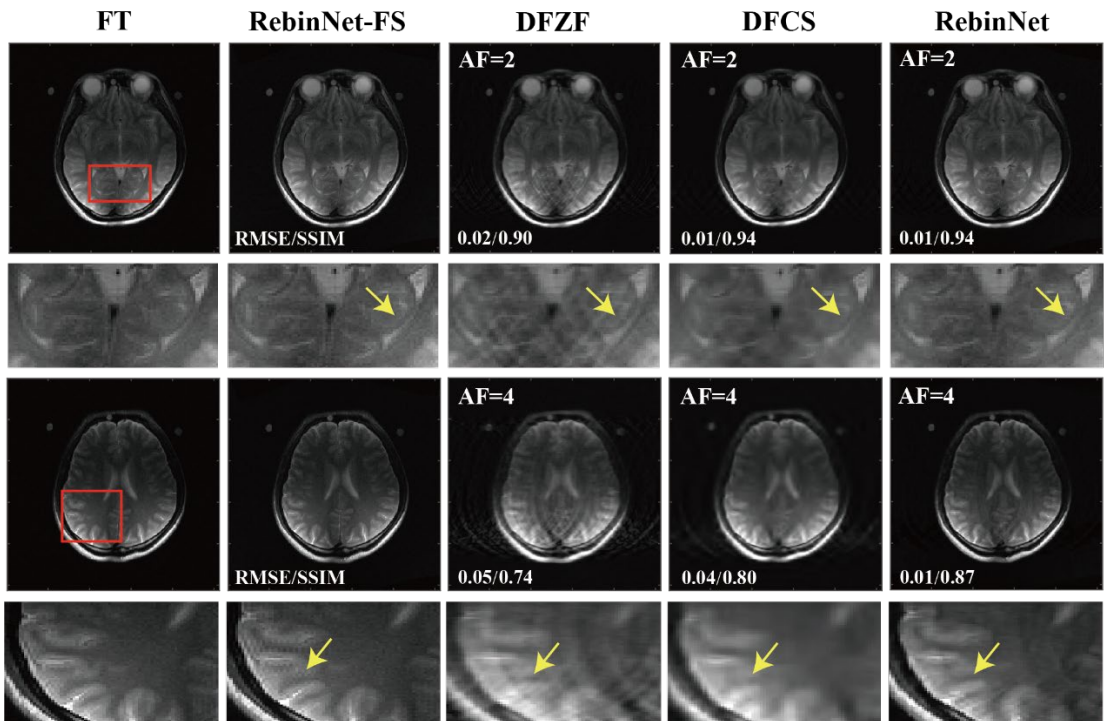

Fig. 8. Volunteer brain images acquired from our MRI-Linac system. The fully sampled reconstructions for FT and RebinNet-FS methods are shown on the left two columns. The DFZF, DFCS and RebinNet were used for image reconstructions at AF=2 (the top row) and AF=4 (the bottom row). The top and bottom slices were acquired at y=-12.5mm and y=13mm, respectively.

## IV. Discussion

The fast acquisition and reconstruction of geometrically accurate images is the ongoing challenge for clinical translations of the MRgRT technique [42-44]. Although MR acceleration techniques with undersampled k-space have shown great potentials to reduce MR acquisition time, the reconstruction methods based on conventional iterative regularization algorithms are still computationally expensive. Here, we leveraged the advance of interpretable unrolled networks to develop a fast image reconstruction pipeline that can reconstruct $B_0$ inhomogeneity distortion-free images directly from undersampled k-space data. The latency of the proposed method was within 0.5s, which is considered acceptable for online and fast image reconstruction during MRI-guided radiation treatments [45, 46]. In addition, the residual geometric distortions after using the RebinNet was less than 2mm. Studies have shown that geometric inaccuracy within 2 mm could result in ⩽5% dosimetry errors and thus is tolerable for accurate absorbed dose delivery [47].

The image domain UnUNet method resulted almost comparable image reconstructions to the RebinNet for the evaluations on simulated brain images. When tested on phantom images that were never seen in the training process, the RebinNet reconstructed better image quality than the UnUNet method, indicating superior generalization ability. The GNL causes sequence-independent geometric distortions; whereas $B_0$ inhomogeneity leads to sequence-dependent distortions which will be affected by the sequence parameters (e.g., bandwidth, applied gradient strength and so on). The GNL distortions were separated by the reversed gradient technique and were corrected using our previously developed DCReconNet method [33] in this work. The performance of distortion-free image reconstruction relies partially on the accurate $B_0$ field characterization [48]. Spherical harmonics and phantom measurements were combined to provide the $B_0$ inhomogeneity information. Since $B_0$ inhomogeneity is different from scanner to scanner, new phantom measurements are required for other MRI systems.

In this work, the standard Cartesian TSE sequences were used to acquire the testing brain and phantom data. Recently, non-Cartesian sequences such as radial and spiral sampling have shown promise for fast tumour tracking in MRI-guided radiotherapy [49]. However, non-Cartesian sequences need to rapidly switch encoding gradients, which would introduce strong eddy currents and thus cause additional image artifacts. This deep learning method has much potential to solve the image reconstruction problems in non-Cartesian sequences.

## V. Conclusion

In this work, a deep-learning based method was developed to reconstruct distortion-free images from B0 inhomogeneity-corrupted k-space with fully sampled and undersampled acquisitions. Simulation and experimental results demonstrated that the RebinNet could preserve image structural details and dramatically improve the computational efficiency in comparison to conventional regularization algorithms. Additionally, the proposed network has shown better generalization ability than the image-domain UnUNet method. Therefore, the RebinNet shows great potentials for facilitating accurate and fast image guidance in radiotherapy treatments.